# Mode imaging and selection in strongly coupled nanoantennas

Jer-Shing Huang[1,*] (黃哲勳), Johannes Kern[1], Peter Geisler[1], Pia Weinmann[2], Martin Kamp[2], Alfred Forchel[2], Paolo Biagioni[3] & Bert Hecht[1,†]

1. Nano-Optics & Biophotonics Group, Experimentelle Physik 5, Physikalisches Institut, Wilhelm-Conrad-Röntgen-Center for Complex Material Systems, Universität Würzburg, Am Hubland, D-97074 Würzburg, Germany

2. Technische Physik, Experimentelle Physik 5, Physikalisches Institut, Wilhelm-Conrad-Röntgen-Center for Complex Material Systems, Universität Würzburg, Am Hubland, D-97074 Würzburg, Germany

3. CNISM - Dipartimento di Fisica, Politecnico di Milano, Piazza Leonardo da Vinci 32, 20133 Milano, Italy

\* jhuang@physik.uni-wuerzburg.de

† hecht@physik.uni-wuerzburg.de

**The number of eigenmodes in plasmonic nanostructures increases with complexity due to mode hybridization, raising the need for efficient mode characterization and selection. Here we experimentally demonstrate direct imaging and selective excitation of the "bonding" and "antibonding" plasmon mode in symmetric dipole nanoantennas using confocal two-photon photoluminescence mapping. Excitation of a high-quality-factor antibonding resonance manifests itself as a two-lobed pattern instead of the single spot observed for the broad "bonding" resonance, in accordance with numerical simulations. The two-lobed pattern is observed due to the fact that excitation of the antibonding mode is forbidden for symmetric excitation at the feedgap, while concomitantly the mode energy splitting is large enough to suppress excitation of the "bonding" mode. The controlled excitation of modes in strongly coupled plasmonic nanostructures is mandatory for efficient sensors, in coherent control as well as for implementing well-defined functionalities in complex plasmonic devices.**

**Introduction**

Plasmonic nanostructures consisting of particular arrangements of closely spaced resonant particles are of great interest since they offer a variety of eigenmodes that evolve due to mode hybridization [1]. Characterization and well-defined excitation of such eigenmodes is important in order to achieve well-defined functionality in devices [2,3] and to successfully apply techniques of coherent control [4-7]. Nanoantennas consisting of two strongly coupled particles can serve as a model system to study the impact of mode selectivity [6,8,9]. Upon illumination nanoantennas confine and enhance optical fields [10,11] and can therefore be used to tailor the interaction of light with nanomatter [12]. Various applications of nanoantennas have been proposed and experimentally demonstrated, including enhanced single-emitter fluorescence [13-15], enhanced Raman scattering [16,17], near-field polarization engineering [18-20], high-harmonic generation [21,22], as well as applications in integrated optical nanocircuitry [23,24]. The longitudinal resonances of a symmetric dipole antenna can be understood in terms of hybridization of the longitudinal resonances of individual antenna arms, caused by the coupling over the narrow feedgap [25,26]. Such coupling causes a mode splitting into a lower-energy "bonding" mode and a higher-energy "antibonding" mode, respectively (Fig. 1). Limited by the uncertainty of conventional nanofabrication, it is often difficult to fabricate antenna arrays with a reproducible gap size below 20 nm, which is necessary to achieve significant energy splitting between the bonding and the antibonding mode. As a result, the existence of the antibonding antenna resonance has hardly been considered, although it may offer interesting opportunities, such as impedance tunability, a high quality factor due to its weakly radiative nature, the launching of propagating plasmon modes with increased propagation lengths [27] and spatial selectivity of optical field distribution. As sketched in Fig. 1, the energy splitting of the modes in a linear dipole antenna increases with reduced feedgap size, revealing the distance dependent interparticle coupling [8,28-32].

Even in presence of strong coupling, anti-bonding antenna modes are not always observed in white-light scattering experiments since their excitation is symmetry-forbidden for normally-incident plane wave sources [33]. The antibonding antenna mode can, however, be excited if the symmetry of the system is broken either by the shape of the structure [34,35] or by the excitation geometry [6,27,36,37]. For example, Liu et al. [27] show theoretically that a dipole located at the end of a pair of gold bipyramids can excite both the bonding plasmonic mode and the antibonding mode. Yang et al. [37] use polarization-sensitive total internal reflection excitation geometry to obtain the spectral signature of transverse coupled electric dipoles in a nanocrystal dimer. Another way to study the resonance of nanoantennas is to measure their two-photon excited photoluminescence (TPPL), which strongly increases when their local field is enhanced by a resonance. To this aim a tightly focused pulsed laser beam is scanned across nanoantennas to obtain TPPL maps [10,38]. Ghenuche et al. [38] demonstrated direct field mapping by scanning large antennas (total length = 1040 nm, gap = 40 nm) over a tightly focused excitation spot (illumination spot of 350 nm) and recording the TPPL signals. However, in refs. [10] and [38] only the bonding antenna mode is considered to explain the recorded TPPL maps.

Here, we exploit the strong coupling between individual nanoantenna arms over a 16 nm feedgap. Such a small feedgap leads to a sufficiently large energy splitting between the hybridized modes and allows for selective mode excitation with focused laser pulses. We demonstrate, for the first time, that the antibonding mode of single crystalline symmetric dipole nanoantennas, which exhibits a significantly lower near-field intensity enhancement compared to the bonding mode albeit a much higher quality factor, can still dominate the TPPL signal in the case of strong coupling. Its TPPL map shows a node line for symmetric excitation which is not seen in bonding mode TPPL maps and clearly indicates selective excitation of the antibonding antenna mode. Our experiments very well match numerical simulations that consider TPPL as a function of the excitation position.

**Experiments and simulations**

Arrays of nanoantennas with very small feedgaps are fabricated reproducibly by focused-ion-beam (FIB) milling of single-crystalline gold platelets placed on top of a cover glass coated with a thin layer of ITO (100 nm) to provide DC conductivity. Since the crystal domain of the self-assembled gold flakes is well defined [39,40], we improve the structuring precision and minimize the plasmon scattering due to the absence of randomly oriented crystal grains [41], normally seen in vapor-deposited multi-crystalline metal layers. In addition, the near-field enhancement of nanoantennas is also improved since we avoid using any adhesion layers, which can change the optical properties drastically [42-45]. The nominal total length of the antennas is varied between 196 nm and 396 nm in steps of 20 nm. Thereby the antenna resonance can be studied with fixed excitation wavelength without any influence of a varying dielectric function. The height, width and the gap size of the antennas are 30 nm, 50 nm and 16 nm, respectively, as measured from scanning electron microscope (SEM) images with an uncertainty of 5 nm. The distance between adjacent antennas is larger than 700 nm to minimize crosstalk and to ensure that only one antenna is illuminated by the focal spot (FWHM = 350 nm), which is slightly smaller than the longest antenna in the array. The ultrashort pulses from a mode-locked Ti:sapphire laser (center wavelength = 828 nm, 80 fs, 80 MHz, average power: 50 µW, Time-Bandwidth Products, Tiger) are coupled into 1.5 m of optical fiber to get stretched to 1 ps, which avoids possible damage of the nanoantennas without decreasing the TPPL signal [46]. The linear polarized fiber output is then collimated, passes a dichroic mirror (DCXR770, Chroma Technology Inc.) and is focused through the cover glass onto the antenna array using an oil immersion microscope objective (Plan-APO 100x, Oil, NA = 1.4, Nikon). The direction of linear polarization of the beam is adjusted using a λ/2-plate. The photoluminescence signal is collected by the same objective and is reflected by the dichroic mirror. Laser scattering and possible second harmonic signals [21,22] are rejected by a holographic notch filter (O.D. > 6.0 at 830 nm, Kaiser Optical System, Inc.) and a bandpass filter (transmission window: 450-750 nm, D600/300, Chroma Technology Inc.) in front of the photon detector (SPCM-AQR 14, Perkin-Elmer). The quadratic dependence on the excitation power confirms that the visible TPPL of gold [47] dominates the recorded antenna emission, while the dependence on the excitation polarization verifies that the longitudinal resonance is excited (see Supplementary Information). To support the experimental results, numerical simulations adopting the nominal antenna dimensions used in FIB milling are performed with a commercial finite-difference time-domain solver (FDTD Solutions v6.5.8, Lumerical Solutions, Inc.). The dielectric constant of gold is modeled according to ref. 48 and ref. 49. The minimal mesh size is set to 1 nm$^3$ yielding a sufficiently high accuracy (see Supplementary Information for simulation details).

**Results and discussion**

We first study the dependence of TPPL efficiency on the antenna length. Figure 2a shows a TPPL map of an antenna array for longitudinally-polarized excitation, along with SEM images of the respective antennas. A dependence of the TPPL signal on the overall antenna length is clearly observed. The TPPL signal first increases and reaches a maximum for antenna 3 (nominal length = 236 nm) and then gradually decays as the antenna length further increases. However, a close look reveals that this trend is not followed by antenna 9 (nominal length = 365 nm) for which the TPPL signal again increases. Furthermore, as the antenna length increases the spot shape obviously changes from slightly elongated Gaussian spots to more elongated ones, and finally divides into two separate spots as highlighted in Fig. 2b-d.

To model the evolution of the TPPL signal and the spot shape transformation with increasing antenna length, we perform numerical simulations for each antenna in the array. Specifically, we displace the exciting Gaussian beam along the antenna axis by 100 nm from the antenna feedgap and record the resulting near-field intensity spectra 5 nm away from the antenna ends. Fig. 3 displays the calculated near-field intensity spectra, which have been normalized with the source spectrum used in the simulation in order to reveal the impulse response [5], i.e. the source-independent antenna response. With a focus diameter comparable to the antenna's total length, symmetry breaking can be achieved whenever the focal spot is longitudinally displaced from the antenna's perpendicular symmetry axis and the dipole forbidden antibonding mode can be excited. Due to the fact that the resonance of individual arms are 180 degree out of phase, the antibonding mode couples much less to the far field and therefore its spectral signature has a much smaller FWHM than the bonding mode resonance. From the simulated spectra, we obtain quality factors of the bonding and antibonding mode of 6.5 and 30.8, respectively, for antenna 3 (see Supplementary Information). Owing to the fact that longer antenna arms exhibit weaker restoring force, the longitudinal resonances of individual antenna arms as well as both hybridized modes red shift with increasing antenna length. As a result, the bonding and the antibonding modes move sequentially in and out of resonance with the fixed laser excitation spectrum (orange shaded area in Fig. 3), which explains the observed TPPL intensity evolution and the intensity recovery for antenna 9 in Fig. 2a. As can be seen in Fig. 3, the simulated bonding resonance peak of antenna 3 (236 nm, black solid line) hits the center frequency of the source. Therefore, maximal excitation efficiency for the bonding mode is expected to be observed for this antenna, which agrees with the experimental observation (Fig. 2a). Same trend of intensity evolution is found for an array of antennas with larger width (70 nm, see Fig. S1), which confirms our model and further demonstrates the high degree of structural control achieved in this study. The simulated spectra in Fig. 3 also reveal that even for a gap as small as 16 nm, the two hybridized modes are not fully separated. This opens the possibility to either excite both modes simultaneously or to selectively excite only one of them. In fact, the intensity recovery observed for longer antennas is a direct evidence of the selective excitation of the antibonding mode.

Besides the intensity variation, the spot shape transformation with increasing antenna length is the other important feature observed in the TPPL map (Fig. 2a). Due to its symmetry, the bonding mode can most efficiently be excited with the focal spot centered at the feedgap. This agrees with the observation of a single bright spot for antenna 3 (Fig. 2b), whose bonding resonance shows maximal overlap with the source spectrum. Antennas with shorter length (<236 nm) also show single spots but decreasing TPPL signal since the bonding resonance blue-shifts away from the excitation. For antennas of intermediate length (between 236 nm and 356 nm), the source spectrally overlaps with both modes. While the excitation beam focused at the feed gap can only excite the bonding mode, a displaced focus is able to excite both hybridized modes. This behavior is reflected by the appearance of a single, but elongated spot in the TPPL map shown in Fig. 2c, resulting from the superposition of a single, centered bonding peak and two displaced antibonding peaks. Considering that the excitation efficiency of the modes is determined by the spectral and spatial overlap with the source as well as the symmetry breaking of the excitation, we note that it is possible to tune the relative amplitudes of bonding and antibonding modes by displacing the laser focus along the antenna's long axis. For very long antennas (>356 nm) the two hybridized modes red shift so much that the source only excites the antibonding mode. Since this excitation is symmetry-forbidden for illumination centered at the feedgap, the result is a near-zero TPPL signal at the antenna gap and therefore the appearance of a two-lobed spot as displayed in Fig. 2d.

To confirm our understanding of the observed TPPL maps, we perform simulations in which we scan the excitation spot over the long antenna axis in order to mimic a line profile recorded in the experiment. The near-field intensity ($|E|^2$) distribution inside the gold arms is recorded for the source center wavelength (828 nm). Based on the fact that TPPL is an incoherent process and the signal depends quadratically on the local field intensity [47], we obtain a quantity that is proportional to the TPPL signal by integrating the square of the near-field intensity ($|E|^4$) [38] over the volume occupied by gold. In Fig. 4a-c, normalized TPPL intensities obtained from simulations are plotted as a function of the source displacement from the feedgap together with experimental data of respective antennas shown in Fig. 2b-d. The simulated line profiles (red line) well reproduce the experimental spot shapes (black dots, raw data) for all three situations (single spot, elongated spot, double spot, respectively) without any free parameters apart from amplitude scaling. Integrating the simulated profiles, we then obtain a quantity that is proportional to the total spatially integrated TPPL signal emitted by each antenna, which is plotted in Fig. 4d together with the corresponding experimental quantity obtained by integrating the photon counts in the bright spots of the experimental TPPL maps (Fig. 2a).

To obtain results with statistical significance we have fabricated four antennas for each nominal dimension. The error bars shown in Fig. 4d indicate the standard deviation of the antenna length and the integrated TPPL signal. The simulated integrated TPPL intensity fits the experimental data in terms of relative amplitude and only slightly underestimates the width of the bonding resonance possibly due to the idealized particle shape used in the simulations. We therefore exclude that the TPPL brightness recovery seen for antenna lengths between 300 nm and 350 nm can be ascribed to a higher-order bonding resonance of the dipole antenna, and conclude that it is due to the excitation of the antibonding mode.

As a further evidence to support our analysis, in Fig. 5 we compare the TPPL maps obtained for two dipole antennas of same length but different width. The observed transition from a single spot to a two-lobed image pattern for these two equally long antennas is caused by decreasing the width of a resonant antenna, which increases the aspect ratio and thus red-shifts the spectrum. This observation further confirms that the two-lobed pattern observed in the TPPL map is due to the excitation of antibonding mode and not a result of any possible excitation via the discontinuity at the antenna edges [38, 50].

**Conclusions**
We have demonstrated the direct observation and selective excitation of the antibonding mode in strongly coupled linear dipole nanoantennas. Excitation of the antibonding mode manifests itself in the observation of a two-lobed pattern in the TPPL map, for which the TPPL intensity minimum occurs at the antenna feed gap, indicating minimum excitation efficiency due to the symmetry-forbidden excitation of the antibonding resonance. In addition to a larger near-field intensity enhancement factor, a smaller feed gap results in a strong coupling and facilitates selective excitation of the different hybridized modes. Due to the change in resonance frequency, field distribution and resulting impedance, as well as its higher quality factor compared to the bonding mode, the antibonding mode can be of great interest for sensor applications, near-field engineering, as well as for applications in plasmonic nanocircuitry since the modes that are launched using the antibonding antenna resonance will have different field distribution and more favorable propagation properties. Furthermore, characterization and selective excitation of modes is a prerequisite for the implementation of coherent control schemes in complex plasmonic nanostructures.

**Acknowledgement**
The authors thank J. Prangsma, S. Großman, C. Brüning, A. Reiserer, C. Rewitz, P. Tuchscherer and T. Brixner for valuable discussions and M. Emmerling for technical support of EBL.

**Author contributions**
J.S.H., P.B., and B.H. conceived the concept of the study. J.S.H., P.W. and M.K. prepared the sample. J.S.H., J.K. and P.G. performed the optical experiments. J.S.H. designed and implemented the simulation; analyzed and assembled the data. J.S.H., J.K., P.G., P.B. and B.H. contributed to the interpretation and critical revision of the article. J.S.H. and B.H. wrote the manuscript. M.K., A.F. and B.H. supervised the study.

**Additional information**
Supplementary information accompanies this paper at www.nature.com/naturephotonics. Reprints and permission information is available online at http://npg.nature.com/reprintsandpermissions/. Correspondence and requests for materials should be addressed to J.S.H. (jhuang@physik.uni-wuerzburg.de) and B.H. (hecht@physik.uni-wuerzburg.de)

# Figure legends

**Fig. 1**: Huang, J. S. *et al.*

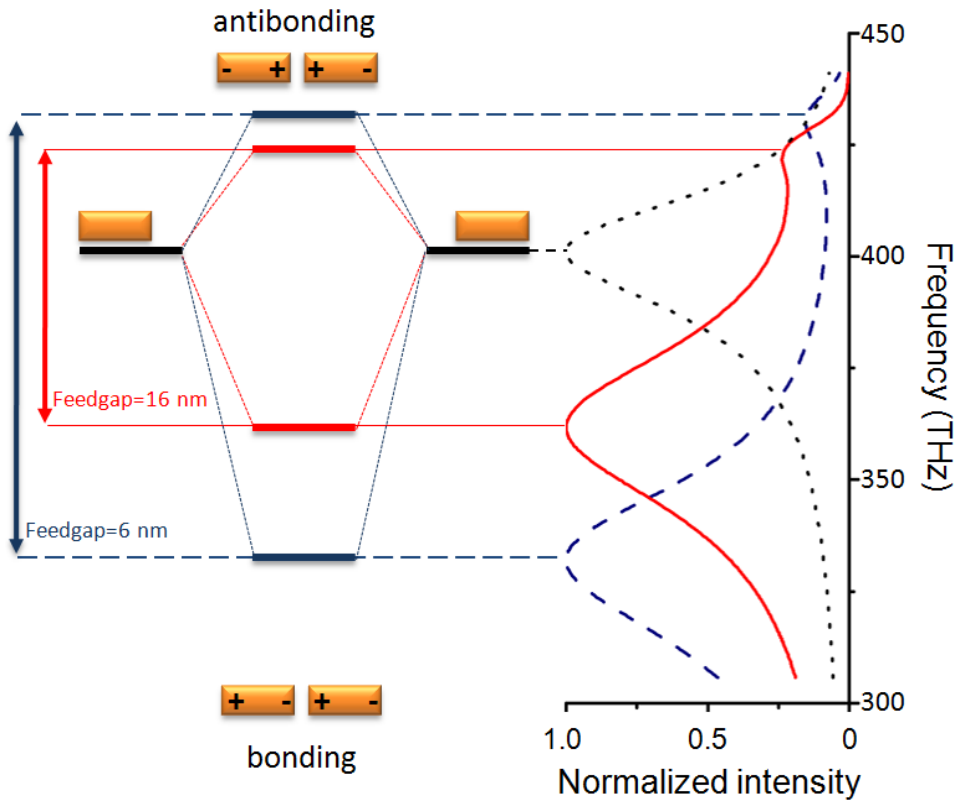

**Figure 1 | Energy-level diagram and simulated near-field intensity spectra**. Near-field intensity spectra are simulated for 30 nm high, 50 nm wide and 110 nm long (arm length) symmetric linear dipole antennas with 6 nm (blue dashed) and 16 nm (red solid) gap, as well as for a single rod with same dimension (black dotted). Spectra are obtained by illuminating the structures with a source displaced 100 nm from the gap centre, by monitoring local fields 5 nm away from the rod (antenna) ends.

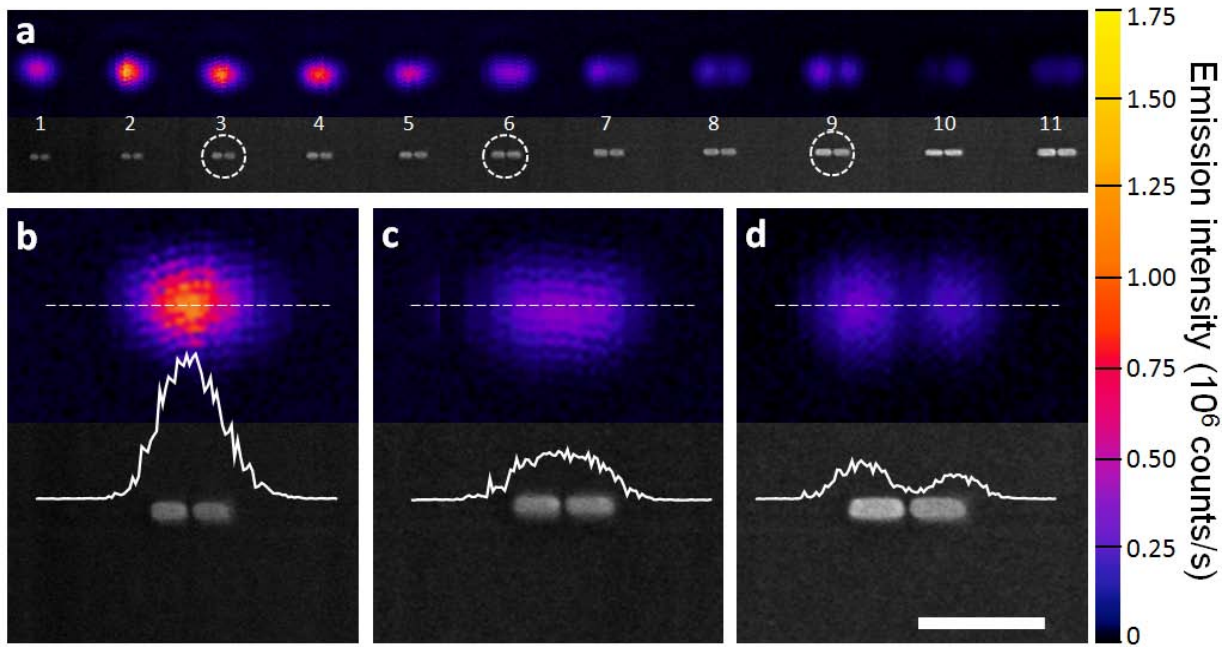

**Figure 2 | Spatially resolved TPPL maps (raw data) and the corresponding SEM images of nanoantennas. a,** TPPL map of an array of antennas with height of 30 nm, width of 50 nm and nominal total length ranging from 216 nm to 396 nm in steps of 20 nm. **b-d,** Zoomed-in TPPL maps of antenna 3, 6, and 9 with nominal total lengths of 236 nm, 296 nm, and 356 nm, respectively, as well as the corresponding SEM images with line-cuts (white dashed line) of the respective TPPL signals along the longitudinal antenna axis. Scale bar, 400 nm.

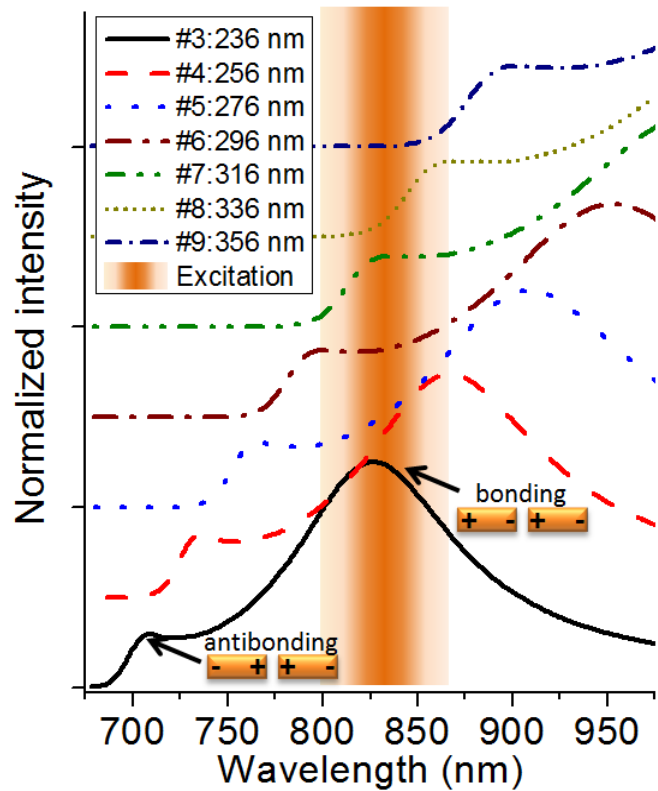

**Figure 3 | Calculated near-field intensity spectra of nanoantennas.** Calculated near-field intensity spectra of antenna 3 to 9 shown in Fig. 2a are plotted with increasing offsets for clarity. The position of the excitation spot is displaced 100 nm from the feedgap along the antenna long axis and the spectra are recorded 5 nm away from the extremities and along the antenna's long axis in order to visualize both modes. With increasing antenna length, the spectrum red-shifts so that the best spectral overlap with the excitation spectrum (orange shaded area) shifts from the bonding to the antibonding mode and beyond.

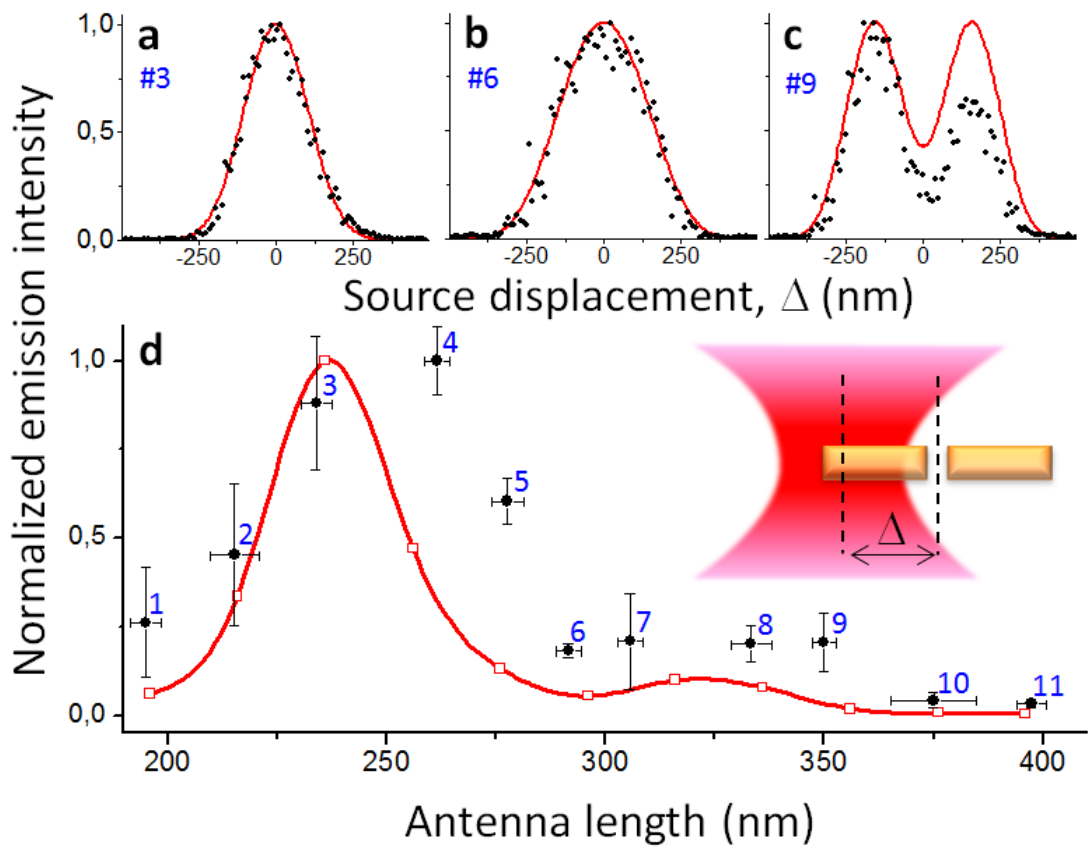

**Figure 4 | One dimensional TPPL profiles and overall brightness with respect to antenna total length. a-c,** Experimental TPPL signals (black dots) plotted with respect to the source displacement, Δ (inset of **d**), show good agreement with simulated line profiles (red solid line) for antenna 3, antenna 6 and antenna 9 with nominal length of 236 nm, 296 nm, and 356 nm, respectively. **d,** The normalized integrated TPPL signals obtained from the simulation (red open squares, red solid line is a guide for the eye) and optical experiment (black dots with error bars) also show fairly good agreement with each other.

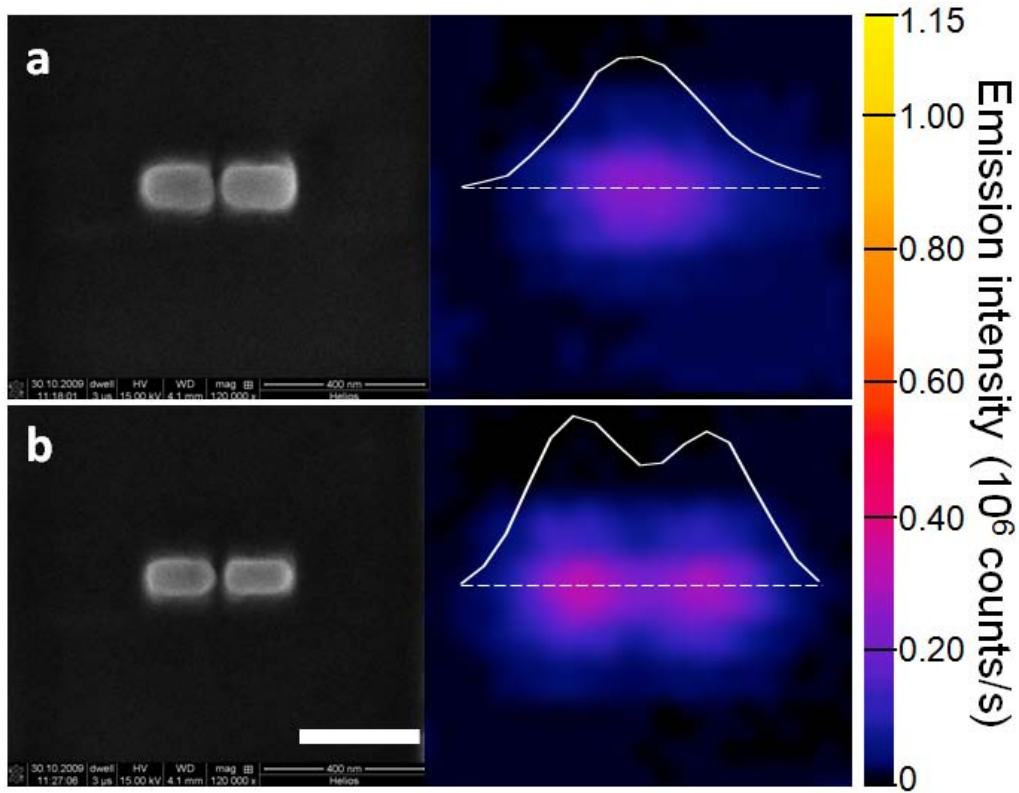

**Figure 5 | SEM images and the corresponding spatially resolved TPPL maps of nanoantennas with same length but different width. a,** SEM image and the corresponding TPPL map together with respective line-cut (white dashed line) emission intensity for a 40 nm high, 383 nm long and 104 nm wide single crystalline nanoantenna with a 20 nm gap. **b,** SEM image and the corresponding TPPL map for a single crystalline nanoantenna with same height, length and gap size with **a** but reduced width (88 nm). The aspect ratios for the antenna in **a** and **b** are 3.7 and 4.3, respectively. The TPPL map changes from a single spot in **a** to a two-lobed pattern in **b** indicating predominant excitation of the antibonding mode. Scale bar, 300 nm.

# Supplementary Information

Mode imaging and selection in strongly coupled nanoantennas

J. S. Huang[1,*](黃哲勳), J. Kern[1], P. Geisler[1], P. Weimann[2], M. Kamp[2], A. Forchel[2], P. Biagioni[3] & B. Hecht[1,†]

1. Nano-Optics & Biophotonics Group, Experimentelle Physik 5, Physikalisches Institut, Wilhelm-Conrad-Röntgen-Center for Complex Material Systems, Universität Würzburg, Am Hubland, D-97074 Würzburg, Germany

2. Technische Physik, Experimentelle Physik 5, Physikalisches Institut, Wilhelm-Conrad-Röntgen-Center for Complex Material Systems, Universität Würzburg, Am Hubland, D-97074 Würzburg, Germany

3. CNISM - Dipartimento di Fisica, Politecnico di Milano, Piazza Leonardo da Vinci 32, 20133 Milano, Italy

* jhuang@physik.uni-wuerzburg.de

† hecht@physik.uni-wuerzburg.de

# 1. Polarization dependence of the antenna resonance

Fig. S1 shows the dependence of the antenna emission on the excitation polarization. To study the mode hybridization due to the strong coupling of longitudinal resonances in the antenna arms, the in-plane excitation polarization is kept parallel to the antenna long axis, i. e. Θ = 0° as defined in Fig. S1a. We have fabricated single-crystalline nanoantenna arrays with arm widths of 50 nm (dashed rectangle, Fig. S1b) and 70 nm (dotted rectangle, Fig. S1b). All arrays show clear dependence of the TPPL intensity on the orientation of the linear excitation polarization as well as the appearance of the two-lobed pattern for the antibonding mode as discussed in the main text. Experimental results presented in the main text are extracted from antenna arrays in the area marked with the dashed rectangle in Fig. S1b.

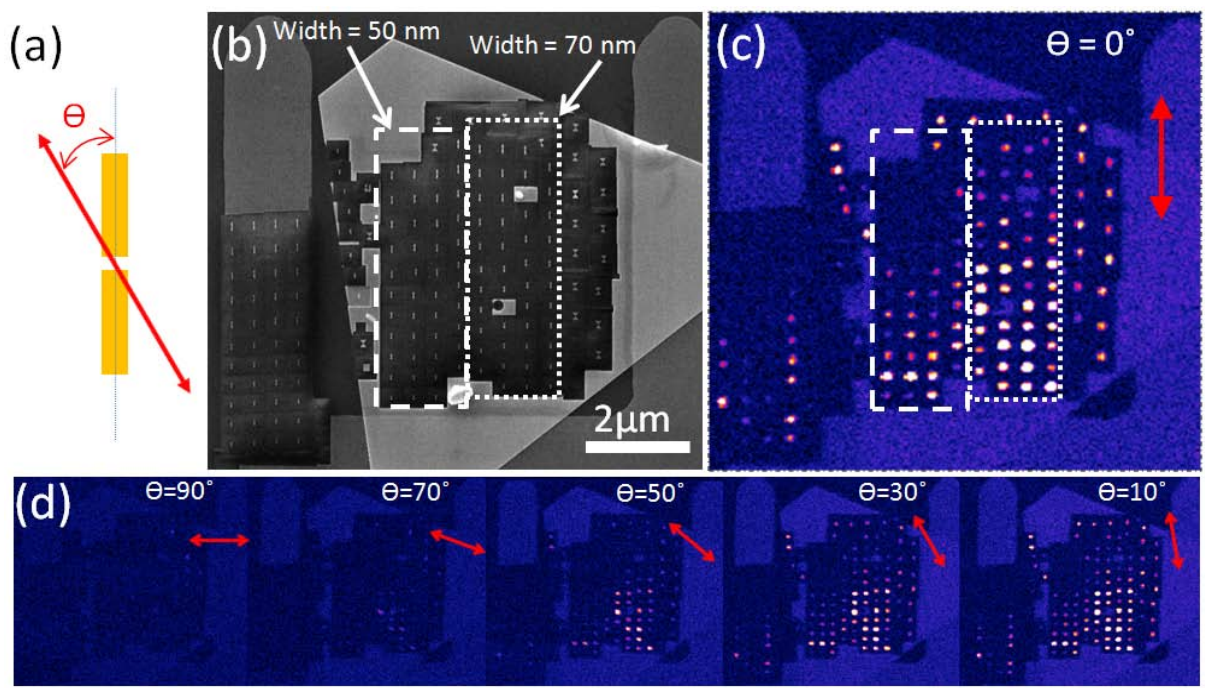

*Figure S1 |* *SEM image and TPPL maps of the whole antenna array obtained with different excitation polarizations. (a) schematic diagram of the linear excitation polarization (red double arrows). Θ is defined as the angle between the excitation polarization and the antenna's long axis; (b) SEM image of the fabricated area including arrays of antennas with nominal width of 50 nm (dashed rectangle) and*
*70 nm (dotted rectangle) ; (c) TPPL map of the corresponding area shown in (b) with longitudinal excitation (Θ = 0°); (d) TPPL maps for various excitation polarizations. Same intensity scale for all TPPL maps.*

# 2. Power dependence of antenna emission intensity

To make sure that the recorded signal is visible TPPL, we have measured the emission intensity with respect to excitation power for different antennas as well as for the unstructured gold area and bare ITO. As shown in Fig. S2, emission signals from both antennas excited at the bonding resonance (marked as "a" with a circle) and antibonding resonance (marked as "b" with a circle) show quadratic dependence on the excitation power while the unstructured gold film (marked as "c" and "d" with circles), as well as bare ITO (marked as "e" with a circle) show much weaker signals with linear dependence due to small leakage of the direct scattering through the filters. Having filtered out the SHG of the gold with a bandpass filter, we assign the quadratic dependence of the antenna emission solely to visible TPPL of gold.

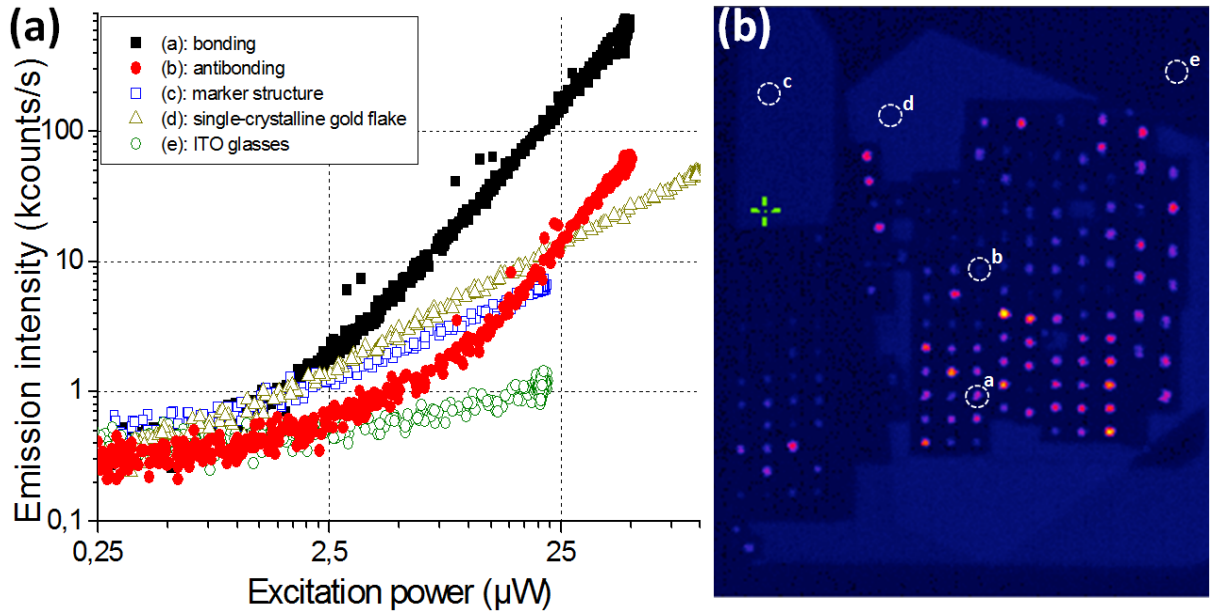

**Figure S2** | *(a) Emission intensity as a function of the excitation power obtained from the area marked with the dashed circles in the emission map (b). The emission signals from antennas with bonding (black solid squares) and antibonding (red solid dots) resonance show quadratic dependence on the excitation power while the scattering from the multi-crystalline gold marker structure (blue open square), single-crystalline gold flake (dark yellow open triangle) and bare ITO glass area (green open circle) show very weak scattering with linear dependence on the excitation power.*

## 3. Simulations parameters

Nominal dimensions used in the FIB milling are adopted in the simulation. Antennas are made of gold placed on top of an ITO layer (thickness?). The dielectric function of gold is described by an analytical model [ref. 46] which fits the experimental data [ref. 45], while the dielectric function of the sputtered ITO layer is based on experimental data [ref. S1]. A multi-coefficient model [ref. S2] is then used to fit the dielectric function within the frequency window of interest to gain speed in the simulation. A uniform mesh volume with discretization of 1 $nm^3$ covers the whole antenna and all the boundaries of the simulation box are set to be at least 700 nm away from the antenna to avoid spurious absorption of the antenna near fields. Since the antenna response is sensitive to local index of refraction, the geometry changes of the substrate due to the FIB milling are taken into account. Figure S3 shows a cross section of the simulated structure. To mimic the experimental conditions, the source is set to have the same spectral width as the laser used in the experiment (821 - 835 nm) and is focused onto the gold/ITO interface using a 1.4 N.A. thin lens approximated by a superposition of 200 plane waves. The resulting focal spot with Gaussian shape (centered in y-direction) is either kept 100 nm displaced from the antenna feedgap or is scanned over the structure in x direction. In order to visualize both hybridized modes, the spectra are recorded 5 nm away from the extremities but on the long axis of the antenna, as indicated by the red cross in Fig. S3. The total electric field intensity ($|Ex|^2+|Ey|^2+|Ez|^2$) distribution inside both antenna arms is recorded using 3D field profile monitors. A quantity proportional to the TPPL signal is obtained by integrating the square of the field intensity ($\iiint (|E_x|^2 + |E_y|^2 + |E_z|^2)^2 dxdydz$) over the volume of the antenna arms.

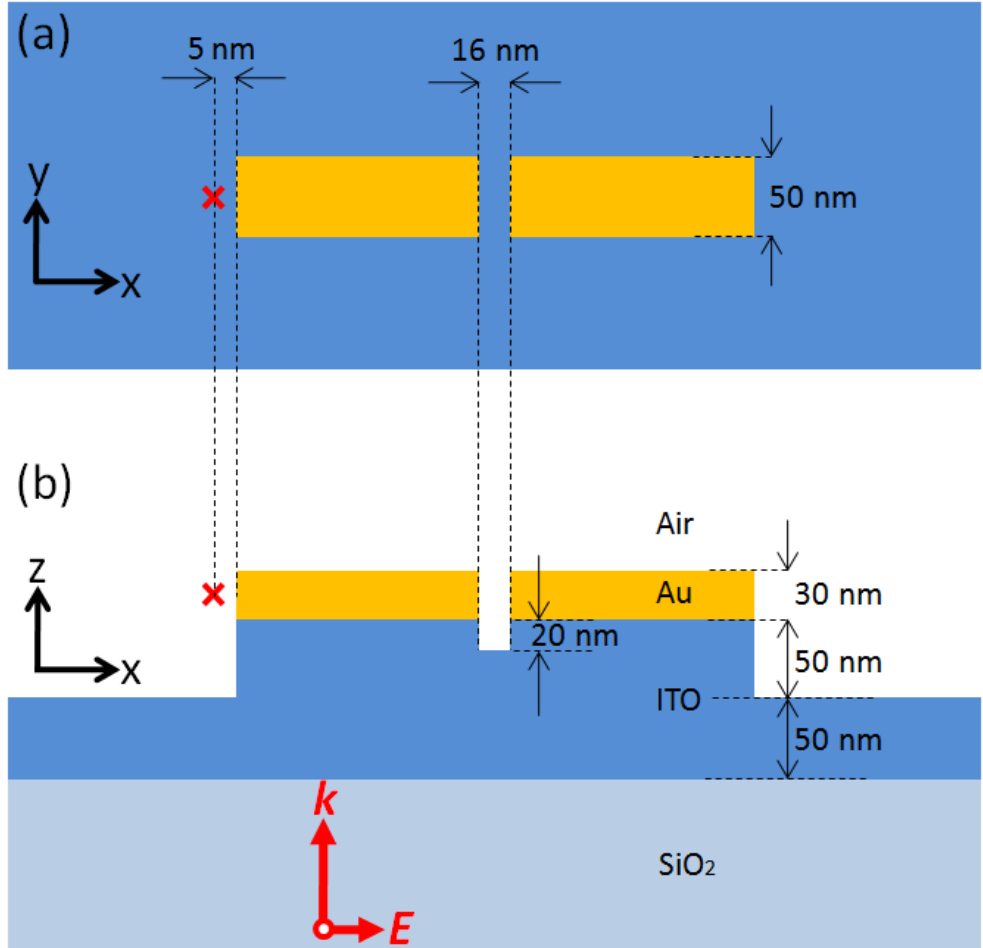

**Figure S3 |** *(a) Top view and (b) cross section of the simulated structure. The excitation source (red arrows) is either fixed with 100 nm displacement from the gap center or scanned over the antenna long axis. Near-field intensity spectra are recorded at the position on the antenna long axis indicated by the red cross. The partial removal of ITO layer due to FIB milling is included in the simulation in order to take the effect of local refractive index variation into account.*

## 4.Quality factor of the bonding and anti-bonding mode

As an example, the quality factor of the bonding and antibonding resonance of antenna 3 (total length = 236 nm) is calculated using $Q = \frac{\lambda}{\Delta\lambda}$, where $\lambda$ is the peak wavelength and $\Delta\lambda$ (in the equation it seems to be lambda prime)is the full width at half maximum. In order to record both the bonding and the antibonding resonance with equal efficiency the source displacement has been optimized to 200 nm from the feedgap. From the spectrum shown in Fig. S4, we obtain quality factors of 31 and 7 for the antibonding and bonding mode, respectively.

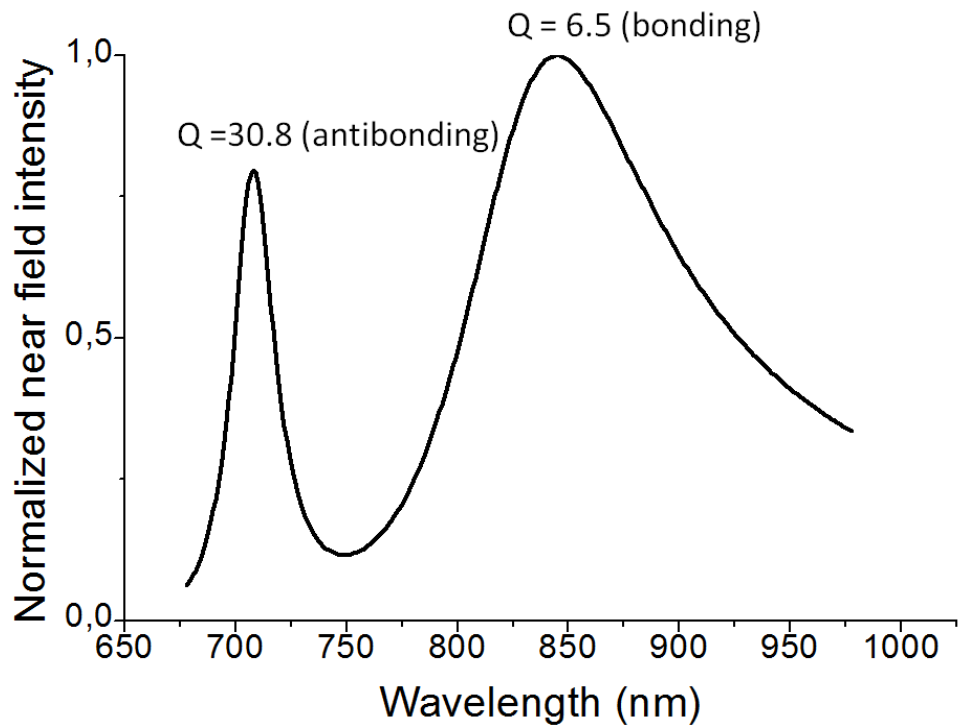

*Figure S4 | Simulated near-field intensity spectrum (open square) of antenna 3 (total length = 236 nm) with 200 nm source displacement to maximize the excitation of the antibonding mode. The quality factor for the bonding and the antibonding mode are 6.5 and 30.8, respectively.*

**References:**

S1. Laux, S. *et al.* Room-temperature deposition of indium tin oxide thin films with plasma ion-assisted evaporation. *Thin Solid Films* **335**, 1-5 (1998).

S2. http://www.lumerical.com/fdtd_multicoefficient_material_modeling.php